\RequirePackage{lineno}
\documentclass[aps,prl,twocolumn,superscriptaddress]{revtex4}

\usepackage{graphicx}
\usepackage{epstopdf}
\usepackage{bm}
\usepackage{amsmath}
\usepackage{amssymb}
\usepackage{hyperref}
\usepackage{float}
\usepackage{array}

\begin{document}

%\title{Role of four-phonon scattering on lattice thermal conductivity in III-V semiconductors: A crucial revisit}
\title{Four-phonon scattering diminishes the optical phonon contribution and isotope effect to thermal conductivity of III-V semiconductors}

\author{Xiaolong Yang}
\affiliation{School of Mechanical Engineering and the Birck Nanotechnology Center, Purdue University, West Lafayette, Indiana 47907-2088, USA.}
\affiliation{Institute for Advanced Study, Shenzhen University, Nanhai Avenue 3688, Shenzhen 518060, China}
\affiliation{Frontier Institute of Science and Technology, and State Key Laboratory for Mechanical Behavior of Materials, Xi'an Jiaotong University, Xi'an 710049, P. R. China.}

\author{Tianli Feng}
\affiliation{School of Mechanical Engineering and the Birck Nanotechnology Center, Purdue University, West Lafayette, Indiana 47907-2088, USA.}
\affiliation{Department of Physics and Astronomy and Department of Electrical Engineering and Computer Science, Vanderbilt University, Nashville, Tennessee 37235, USA.}
\affiliation{Materials Science and Technology Division, Oak Ridge National Laboratory, Oak Ridge, Tennessee 37831, USA. }

\author{Ju Li} \affiliation{Department of Nuclear Science and
  Engineering and Department of Materials Science and Engineering,
  Massachusetts Institute of Technology, Cambridge, Massachusetts
  02139, USA.}

\author{Xiulin Ruan}
\email{ruan@purdue.edu}
\affiliation{School of Mechanical Engineering and the Birck Nanotechnology Center, Purdue University, West Lafayette, Indiana 47907-2088, USA.}

\begin{abstract}
Recent studies reveal that four-phonon scattering is generally important in determining thermal conductivities of solids. However, these studies have been focused on materials where thermal conductivity $\kappa$ is dominated by acoustic phonons, and the impact of four phonon scattering, although significant, is still generally smaller than three-phonon scattering. In this work, taking AlSb as example, we demonstrated that four-phonon scattering is even more critical to three-phonon scattering as it diminishes optical phonon thermal transport, and therefore significantly reduces the thermal conductivities of materials in which optical branches have long three-phonon lifetimes. Also, our calculations show that four-phonon scattering can play an extremely important role in weakening the isotope effect on $\kappa$. Specifically, four-phonon scattering reduces the room-temperature $\kappa$ of the isotopically pure and natural-occurring AlSb by 70$\%$ and 50$\%$, respectively. The reduction for isotopically pure and natural-occurring c-GaN is about 34$\%$ and 27$\%$, respectively. For isotopically-pure w-GaN, the reduction is about 13$\%$ at room temperature and 25$\%$ at 400 K. These results provided important guidance for experimentalists for achieving high thermal conductivities in III-V compounds for applications in semiconductor industry.

\end{abstract}

\maketitle
%\linenumbers
III-V compound semiconductors including AlSb, BAs, and GaN, have attracted remarkable attention due to their extraordinary electronic, thermal,
and optical behaviors, with great potential for applications in high-power electronics, optoelectronic devices, thermal management, light-emitting diodes, field emission devices, solar cells, and nano generators, etc. In particular, owing to the peculiar phonon band structure such as large acoustic-optical (a-o) band gap, acoustic bunching and the flatness of optical branches, these materials tend to exhibit unusually high thermal conductivity, which thus have been promising candidates for heat dissipation in nano-electronics as microelectronic devices shrinks. To this end, in recent years thermal transport in III-V compounds has been extensively investigated by both experimental and theoretical works. And recently, first-principles calculations based on the solution of the phonon Boltzmann transport equation (BTE) have made significant progress in terms of accurately predicting lattice thermal conductivity of a great number of semiconductor systems \cite{Broido2007,Garg2011,Ward2009,Li2012a,Tian2012,Yang2019,Li2012}. By combining the density functional calculations and the linearized BTE, Lindsay and co-workers have calculated lattice thermal conductivities of cubic aluminum-V, gallium-V, and indium-V compounds based on the lowest anharmonic phonon scattering\cite{Lindsay2013}. They performed anharmonic lattice dynamics (ALD) calculation based on density functional theory (DFT) to predict thermal conductivies of wurtize-GaN and cubic-GaN \cite{Lindsay2012}. With same approach, they also examined the thermal conductivities and isotope effect of cubic III-V boron compounds \cite{Lindsay2013a}. Despite great efforts have been made to study thermal transport in III-V compounds, however, due to ignoring higher-order four-phonon scattering, these previous works usually give much large overestimation of thermal conductivities as compared to experimental measurements at higher temperatures or even room temperature (RT). For instance, the zinc-blende BAs was predicted to have a thermal conductivity $\sim$ 2200 W/m K at RT with three-phonon scattering, an order of magnitude higher than the experimental value\cite{Lindsay2013a}. Also, the previous prediction on thermal conductivity of cubic AlSb within three-phonon theory is as high as 100 W/mK at RT\cite{Lindsay2013}, which is substantially higher than measured value by experiment, $\sim$50 W/mK \cite{Steigmeier1966}.

Only recently we have developed a formalism to explicitly determine four-phonon scattering from perturbation theory \cite{Feng2016}, and also demonstrated that four-phonon processes can actually remedy the discrepancies between the previous calculation and experiment and even play a dominant role in certain materials \cite{Feng2017,Feng2018}. Notably, our prediction on BAs thermal conductivity $\kappa$ has recently been confirmed by experiments
\cite{Tian2018,Li2018,Kang2018}, manifesting the robustness of our state-of-the-art approach to predicting lattice thermal conductivities of semiconductors. Although the research techniques of predicting thermal conductivity of semiconductor materials are gradually maturing, nonetheless, thermal transport properties of III-V compounds such as AlSb, c-GaN, and w-GaN, are still not well understood yet, thus inspiring us to revisit their lattice thermal conductivities and understand how previously neglected four-phonon scattering affects their thermal transport properties, which is of both practical and fundamental importance. Moreover, we notice that a few recent studies \cite{Feng2017,Feng2018,Ravichandran2019}within the framework of four-phonon theory have been mainly focused on materials where $\kappa$ is dominated by acoustic phonons, e.g. BAs, and the impact of four phonon scattering, although significant, is still generally smaller than three-phonon scattering.

Of particular note is that some previous predictions indicate that the optical phonons provide major contributions to $\kappa$ in some III-V semiconductors such as the aluminum-V compounds \cite{Broido2012,Tian2012,Li2012}. A typical example is zinc-blende AlSb: Lindsay et \textit{al}. has demonstrated that optical phonons contribute more than the acoustic phonons to the total $\kappa$ even at RT \cite{Lindsay2013}. This is in sharp contrast to BAs for which acoustic phonons dominate thermal conductivity. However, their prediction was limited to only three-phonon scattering. Hence, it is very worth exploring two open questions in such optical phonon-dominated thermal materials: (1) How much does the four-phonon scattering impact the overall thermal conductivity? (2) How much does the four-phonon scattering impact the relative contributions of optical phonons?

Additionally, as aforementioned above, semiconductors like BAs, AlSb, c-GaN, and w-GaN have the unqiue phonon band features (i.e. the large a-o gap, acoustic bunching, and the flatness of the optical bands, etc.), which make the satisfaction of three-phonon selection rules quite difficult if not impossible. This will weaken the coupling or interaction between acoustic and optical phonons and lead to relatively strong phonon-isotope scattering, thus giving the large isotope effect on $\kappa$ in these materials. For instance, Lindsay et \textit{al}. has predicted with only lowest-order anharmonicity that the isotope effect in w-GaN is very large, and at around 30K the calculated $\kappa$ for isotopically pure GaN is almost 7 times higher than that with naturally occurring isotope concentration \cite{Broido2012}. In essence, the origin of the isotope effect is the interplay between intrinsic anharmonic phonon-phonon scattering and phonon-isotope scattering. Although the three-phonon process is largely restricted by the peculiar phonon structure, higher-order four-phonon scattering, however, cannot be restricted by these band characteristics. Thus, a natural question is as follows: how important does four-phonon scattering in the isotope effect in these materials?

Inspired by the above motivations, in this work, by solving the linearized phonon BTE based on first-principles calculations, we systematically study lattice thermal conductivities and isotope effects in BAs, AlSb, c-GaN, and w-GaN with four-phonon included.
Particularly, we demonstrate that after considering four-phonon scattering in AlSb, the relative contribution of optical phonons to thermal conductivity diminishes. Unprecedentedly, this leads to about 50$\%$ reduction in $\kappa$ of naturally occurring AlSb even at RT, indicating that unlike in BAs, four-phonon scattering plays a dominant role over three-phonon scattering in AlSb. Further phonon scattering analysis shows that the four-phonon processes play a dominant role in the intrinsic resistance for optic phonons, due to highly restrictive selection rules of three-phonon scattering. The interplay of intrinsic phonon-phonon scattering and phonon-isotope scattering is also discussed by highlighting the role of four-phonon scattering in determining the $\kappa$ of these four materials. Also, by comparison with AlSb, BAs, c-GaN, and w-GaN, we demonstrate that the interplay among the large a-o gap, bunching of acoustic branches, and low-frequency optical bands together decides the importance of four-phonon scattering in thermal transport properties of materials.

Within the framework of BTE, the lattice thermal conductivity $\kappa$ along the transport direction can be determined by summing up the contributions from each phonon branch $\alpha$ and integrating over the first Brillouin zone (BZ),

\begin{equation}
\begin{split}
\label{eq:1}
\kappa=\frac{1}{k_{\rm B} T^2}\frac{1}{8\pi^3}\sum\limits_{\alpha}\int\limits_{BZ} f_0\left(\omega_{\alpha,\textbf{q}}\right)\left[f_{\rm 0}\left(\omega_{\alpha,\textbf{q}}\right)+1\right] \\
\upsilon_{\alpha,\textbf{q}}^2\hbar^2\omega_{\alpha,\textbf{q}}^2\tau_{\alpha,\textbf{q}} d\textbf{q},
\end{split}
\end{equation}

\noindent $\upsilon_{\alpha,\rm \textbf{q}}$ the phonon group velocity along the transport direction, $\tau_{\alpha, \rm \textbf{q}}$ the phonon lifetime and $f_{\rm 0}$ denotes phonon occupation number that obeys the Bose-Einstein distribution.

To obtain accurate $\kappa$, we employ an iterative scheme mixing three-phonon scattering to exactly
sove the phonon BTE based on ab-initio calculations\cite{Ward2007,Esfarjani2011,Lindsay2013a},
by considering isotopic scattering and intrinsic anharmoncity up to fourth order.
To effectively reduce the computational cost,
herein four-phonon scattering rates $\tau_4^{-1}$ are included into the iterative scheme within the relaxation time approximation (RTA) level,
similar to implementing phonon-isotope and phonon-boundary scattering terms,
which has been demonstrated to be valid when four-phonon scattering
is dominated by Umklapp processes \cite{Feng2017}.
Solving numerical phonon BTE requires accurate calculations of the harmonic and anharmonic interatomic force constants (IFCs),
which are obtained from density functional theory (DFT) within the local density approximation (LDA)
using the first principle-based VASP package\cite{Kresse1993,Kresse1996,Kresse1996a}.
The harmonic IFCs were calculated using 4$\times$4$\times$4 supercells and 4$\times$4$\times$4 Monkhorst-Pack grid within a finite displacement scheme,
as implemented in the open source software packages Phonopy \cite{Togo2008}.
The non-analytical term was included to capture LO/TO splitting for all the materials studied.
The third-order IFCs were calculated through Thirdorder \cite{Li2014},
a package of ShengBTE, considering up to the fifth nearest neighbor.
The fourth-order IFCs were calculated out to the second nearest neighbors.
The 4$\times$4$\times$4 primitive cells and 4$\times$4$\times$4 Monkhorst-Pack grid are adopted to calculate the 3rd and 4th IFCs from a finite differences method.
With these IFCs, the phonon frequencies and velocities are then obtained by diagonalizing the dynamical matrix,
and the phonon scattering rates are calculated from the iterative solution of the BTE described above.
The thermal conductivity is solved with a $16\times16\times16$ $\textbf{q}$-mesh. In the
cell relaxation, the total energy and Hellmann-Feynman force convergence thresholds are set to $10^{-10}$ eV and $10^{-6}$ $\rm{eV/\AA}$, respectively.
The energy cutoff was determined by adding 30$\%$ to the highest energy cutoff for the pseudopotentials.
The expressions for all involved scattering rates in this work have been given in detail previously \cite{Feng2017,Feng2018}.

\begin{figure*}[hbtp]
	\centerline{\includegraphics[width=11cm]{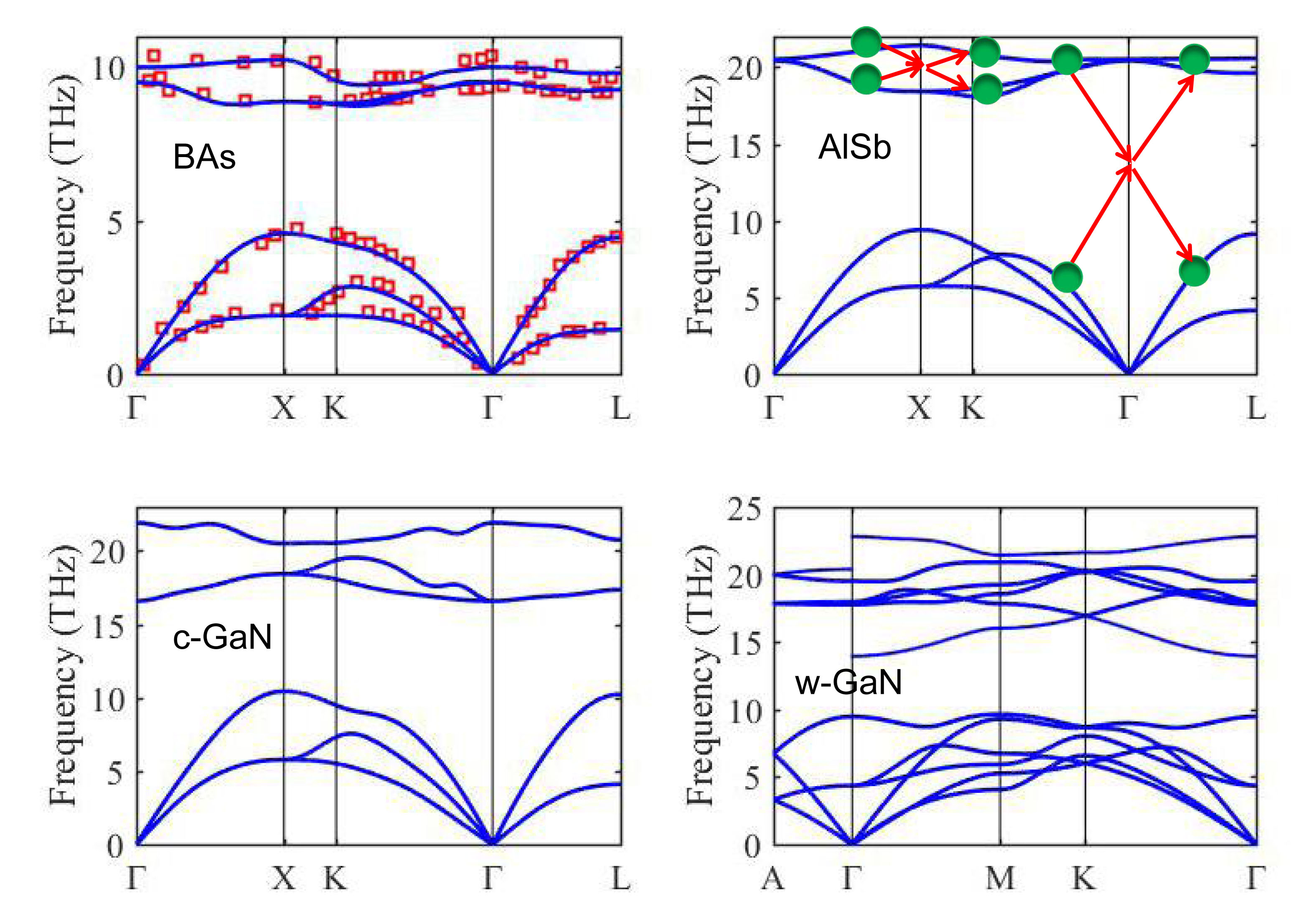}}
	\caption{\label{band}
Phonon dispersions of AlSb, BAs, c-GaN, and w-GaN along high-symmetry directions in the Brillouin Zone (BZ). The solid lines are the results from ab initio calculations in this work. The red squares are experimental data points for AlSb from Ref.\cite{Madelung}. Green balls and red arrows illustrate the recombination channels of four-phonon scattering processes in AlSb.}
\end{figure*}

The calculated phonon dispersions of AlSb, BAs, c-GaN, and w-GaN along the high-symmetry directions are plotted in Fig.\ref{band}. For AlSb, BAs, and c-GaN, the phonon dispersions are very similar along the main high-symmetry $\textbf{q}$-points in the irreducible BZ, and there exists 3 acoustic and 3 optical phonon branches corresponding to the 2 atoms per primitive cell, separated into two parts by the large phonon band gap due to the mass difference between the component elements. For w-GaN, the 12 phonon branches corresponding to the 4 atoms in the primitive cell are also separated into two parts by a large phonon bandgap, of which the upper parts are 6 optical branches while lower ones include 3 acoustic branches and 3 optical branches. In these materials, the large phonon band gap will forbid two acoustic phonons combining into an optical phonon and thus weaken the coupling between acoustic and optical phonons due to the strong restriction of simultaneous energy and momentum conservations. However, it will not be restrictive for four-phonon processes between acoustic and optical phonons. As illustrated in the case of AlSb in Fig.\ref{band}, the recombination processes of four-phonon scattering can easily occur. In addition to the large band gap, we also see that in BAs the three acoustic branches in the frequency range from 4THz to 10THz are close to each other like a bunching, which will suppress the phase space for the three-acoustic-phonon scattering channels (so-called $aaa$ scattering) since the summation of the energies of two acoustic phonons can hardly reach that of the acoustic phonon, as proven previously\cite{Lindsay2013a,Feng2017}. Comparatively, unlike BAs, the three acoustic branches in AlSb are relatively dispersive, implying the stronger $aaa$ scattering phase space in AlSb. Besides, the phonon dispersions of these four materials exhibit the salient flatness of the optical branches, especially for AlSb, which will largely weaken three-optical-phonon scattering processes (so called $ooo$ scattering) due to the energy selective rule. Also, the flatness of optical band reflects that these materials have relatively small group velocities of optical modes. Conventionally, optical phonons should have little contribution to their lattice thermal conductivities. However, what makes AlSb stand out is that based on previous work \cite{Lindsay2013} optical phonons play a dominant contribution to the $\kappa$ over acoustic phonons even at RT. Consequently, we concluded that the large contribution of optical modes to the $\kappa$ in AlSb probably originates not from large velocities, but from the large lifetimes of optical modes, since in AlSb $aao$ and $ooo$ scattering channels are heavily weakened due to the large a-o gap and the flatness of the optical bands, and $aoo$ scattering actually provides the only intrinsic resistance for optical phonons. Hence, four-phonon scattering may become non-negligible for the optical contributions to $\kappa$.

\begin{figure*}[hbtp]
	\centerline{\includegraphics[width=16cm]{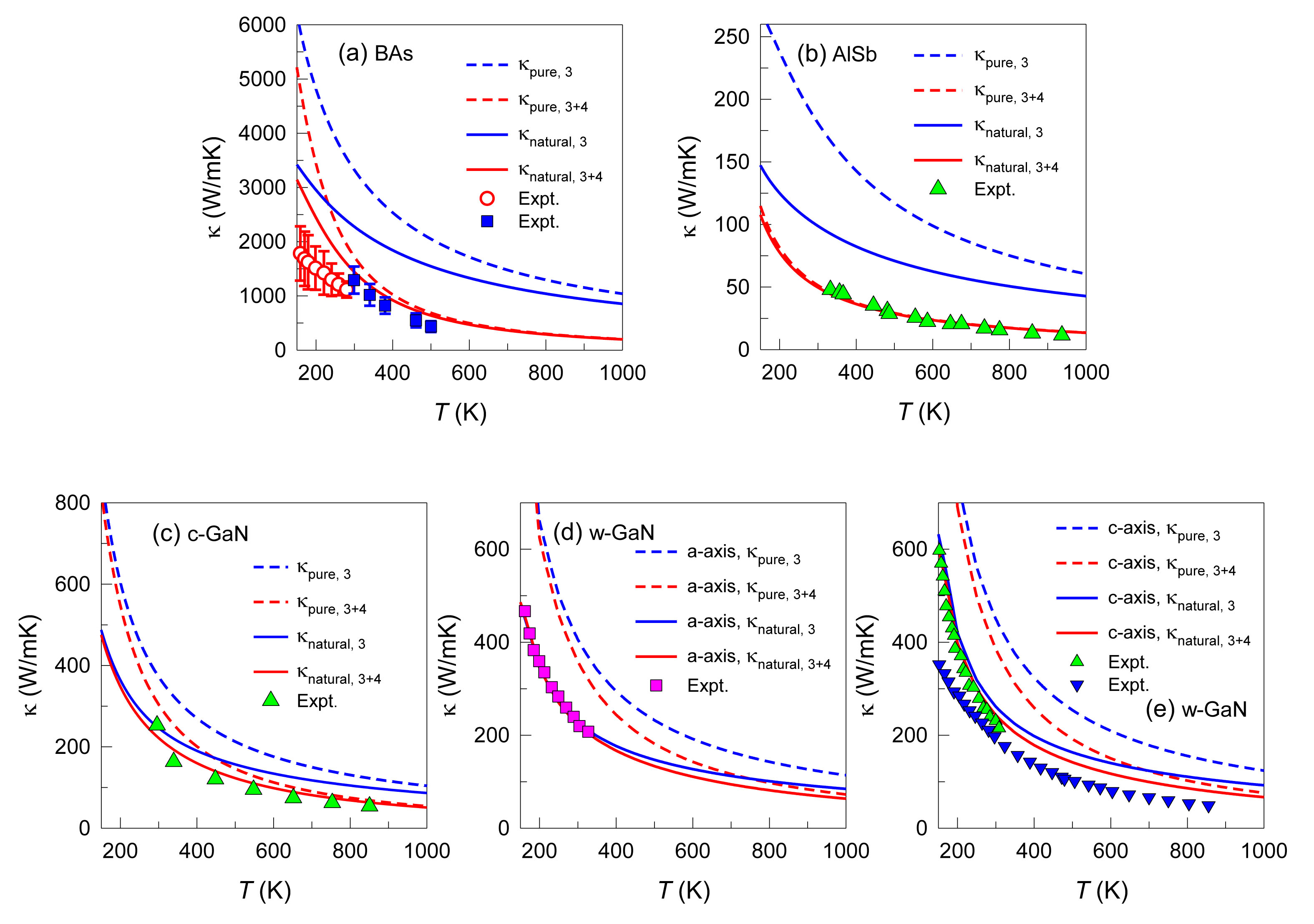}}
	\caption{\label{kappa}
 Lattice thermal conductivity as a function of temperature for BAs (a), AlSb (b), cubic GaN (c), and wurtzite GaN along different directions (d-e).
 Dashed lines represents the calculated $\kappa$ for isotopically pure materials, $\kappa_{\rm pure}$) and solid lines represents the calculated $\kappa$ with naturally occurring materials, $\kappa_{\rm natural}$); The blue lines gives the calculated  $\kappa$ with only three-phonon scattering and red lines give the results after including four-phonon scattering. All symbols represent experimental data for natural occurring materials: blue squares and red circles in BAs are from Ref.\cite{Tian2018}, green triangles in AlSb are from Ref.\cite{Steigmeier1966}, green triangles in c-GaN are from Ref.\cite{Jeowski2003}, pink squres in w-GaN along a axis are Ref.\cite{Jeowski2003}, and green triangles and blue triangles are from Refs.\cite{Jeowski2003,Zheng2019} for w-GaN along c axis.}
\end{figure*}

Our main results are shown in Fig.\ref{kappa}. The calculated $\kappa$ with four-phonon scattering included are compared with those without four-phonon term for naturally occurring BAs, AlSb c-GaN, and w-GaN, as well as corresponding isotopically pure materials, respectively. Blue dashed curves corresponding to $\kappa_{\rm pure,3}$ for isotopically pure materials, and blue solid curves corresponding to $\kappa_{\rm natural,3}$ for naturally occurring materials are calculated results with anharmonicity up to third order; Red dashed curves corresponding to $\kappa_{\rm pure, 3+4}$ for isotopically pure materials, and red solid curves corresponding to $\kappa_{\rm natural, 3+4}$ give the calculated results after including four-phonon scattering; And all symbols represent the measured $\kappa$ of Refs.\cite{Steigmeier1966, Tian2018,Jeowski2003,Zheng2019}. Also, for ease of appreciating the practical significance of four-phonon scattering, our calculated RT values of $\kappa_{\rm pure}$ and $\kappa_{\rm natural}$ for these four materials are also shown in Table \ref{table:kappa}. In terms of naturally occurring materials, it is found that for both BAs and AlSb when anharmonicity is considered up to third order, the values of thermal conductivity $\kappa_{\rm natural, 3}$ are much overestimated over the entire temperature range as compared to the experimental data \cite{Steigmeier1966,Tian2018}. Surprisingly, after including four-phonon scattering, our prediction yields the $\kappa_{\rm natural,3+4}$ values in reasonable agreement with experimental measurement. For BAs when four-phonon scattering is included, calculated RT $\kappa_{\rm natural, 3+4}$ will decrease by 37$\%$, changing from 2276 to 1441 W/mK, which is quantitatively consistent with previous work\cite{Feng2017,Tian2018}. Strikingly, in AlSb we find that four-phonon scattering dramatically lowers the value of RT $\kappa_{\rm natural}$ from 99 to 50 W/mK, by almost 50$\%$, demonstrating the unprecedented significance of four-phonon scattering in AlSb in terms of reducing $\kappa$. 

As is seen from Figs.\ref{kappa}(c-e), for c-GaN and w-GaN, although four-phonon scattering is not as important as that in BAs and AlSb, it still plays a indispensable role in determining $\kappa$, especially at high temperatures. For natural occurring c-GaN and w-GaN, three-phonon predictions can yield the values of $\kappa$ in good accordance with experiment below RT. As $T$ rises, however, obvious deviations from experiment occur. At RT, three-phonon predicted values ($\kappa_{\rm natural, 3}$=246 W/mK and $\kappa_{\rm pure, 3}$=372 W/mK ) are obviously higher than those of four-phonon theory ($\kappa_{\rm natural, 3+4}$=222 W/mK and $\kappa_{\rm pure, 3}$=304 W/mK ) for c-GaN. Similarly, in w-GaN the three-phonon theory gives room-temperature values 
($\kappa_{\rm natural, 3}$=228 W/mK and $\kappa_{\rm pure, 3}$=406 W/mK along the a-axis, $\kappa_{\rm natural, 3}$=261 W/mK and $\kappa_{\rm pure, 3}$=450 W/mK along the c-axis), whereas four-phonon theory presents the lower results with $\kappa_{\rm natural, 3+4}$=223 W/mK and $\kappa_{\rm pure, 3+4}$=359 W/mK along the a-axis, and $\kappa_{\rm natural, 3+4}$=243 W/mK and $\kappa_{\rm pure, 3+4}$=385 W/mK along the c-axis, respectively. At temperature above 400K, our prediction of $\kappa$ including four-phonon scattering deviates significantly from that only involving three-phonon scattering. For c-GaN in particular, at around 850K, three-phonon prediction alone gives $\kappa$ $\sim$ 99 W/mK, much larger than the experimental value, 54 W/mK \cite{Jeowski2003}. After including four-phonon scattering, our prediction can agree well with the available measurements in the whole $T$ range. The disagreement between experiment \cite{Zheng2019} and our theoretical calculation for w-GaN in Fig.\ref{kappa}(e) is mainly ascribed to some defects in experimental samples such as dislocations, impurities, etc. 

Another point worth noting in Fig.\ref{kappa} is that phonon-isotope scattering has a large impact on thermal conductivities for these four materials studied. Comparing $\kappa_{\rm natural}$ and $\kappa_{\rm pure}$ of these materials, we can see that with only third-order anharmonicity considered, phonon-isotope scattering can greatly reduce thermal conductivities, especially for low temperatures. As seen in Table \ref{table:kappa}, AlSb has $\kappa_{\rm pure, 3}$ =181 W/mK around RT being almost two times higher than $\kappa_{\rm natural, 3}$=99 W/mK, the RT $\kappa_{\rm pure, 3}$ for BAs is 3322 W/mK far larger than $\kappa_{\rm natural, 3}$=2276 W/mK, and the RT $\kappa_{\rm pure, 3}$ for c-GaN is 372 W/mK much higher than $\kappa_{\rm natural, 3}$=246 W/mK. For w-GaN, the RT $\kappa_{\rm pure, 3}$ along the a-axis and c-axis are 406 W/mK and 450 W/mK, respectively, being almost two times larger than corresponding $\kappa_{\rm natural, 3}$.  This intriguing finding is due to the interplay between phonon-phonon and phonon-isotope scattering. Since in these materials with large a-o gap and atomic mass difference, the anharmonic three-phonon scattering is relatively weak and phonon-isotope scattering is relatively more important in determining $\kappa$. Remarkably, we find that after considering four-phonon scattering the isotope effects become less important but certainly still cannot be neglected, as inclusion of four-phonon scattering makes the intrinsic anharmonic phonon scattering stronger, thus weakening the relative contribution of phonon-isotope scattering to $\kappa$. We can clearly see from Table \ref{table:kappa} that at RT $\kappa_{\rm pure, 3+4}$=51 W/mK and $\kappa_{\rm natural, 3+4}$=50 W/mK for AlSb, and $\kappa_{\rm pure, 3+4}$=1721 W/mK and $\kappa_{\rm natural, 3+4}$=1441 W/mK for BAs. For both c-GaN and w-GaN, it is quite different from AlSb and BAs that after including four-phonon scattering the values of $\kappa_{\rm pure, 3+4}$ are still much larger than those of
$\kappa_{\rm natural, 3+4}$, as listed in Table \ref{table:kappa}, indicating the strong isotope effect in these two systems.

\begin{figure}[hbtp]
	\centerline{\includegraphics[width=8cm]{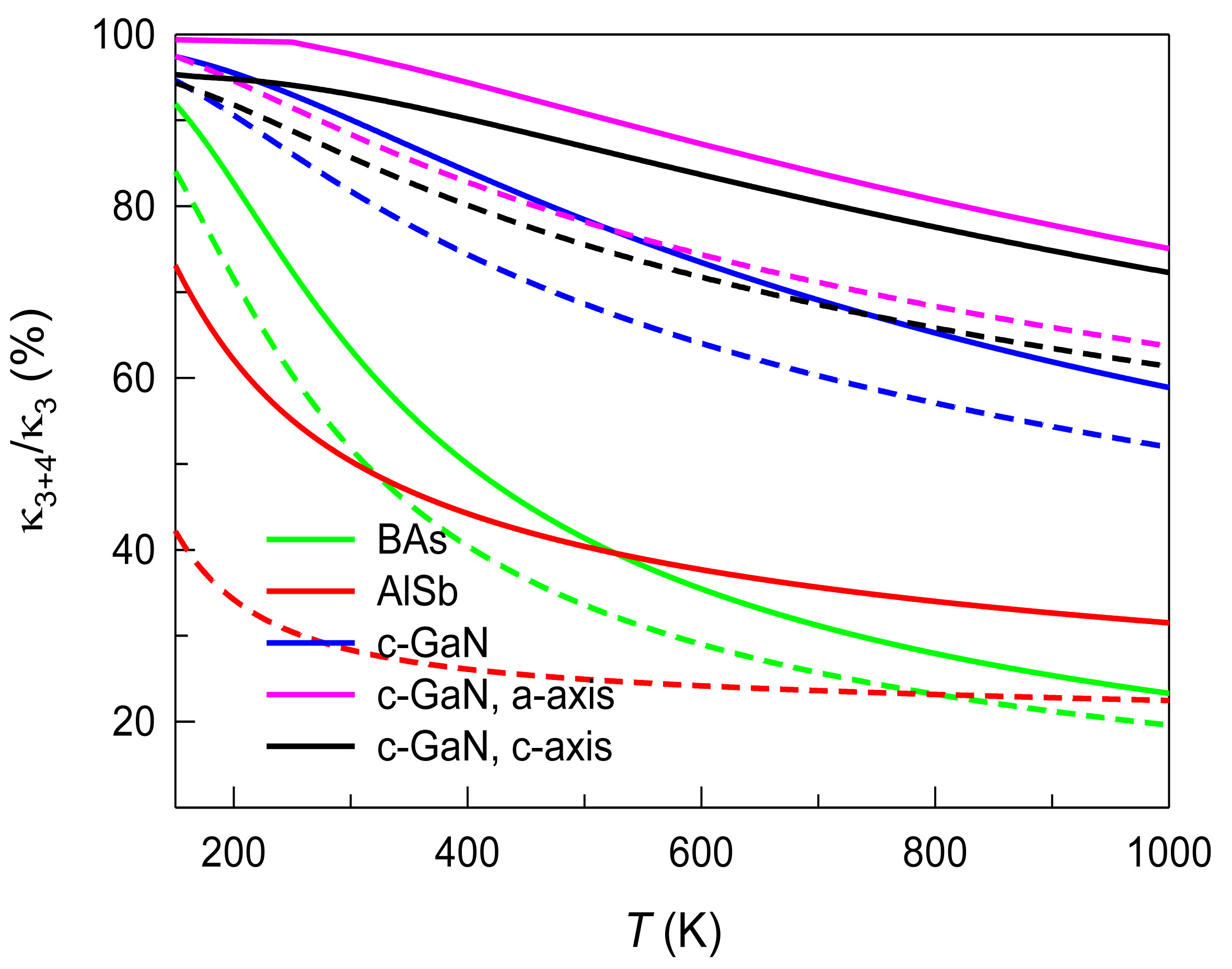}}
	\caption{\label{relative_kappa}
 The relative thermal conductivities $\kappa_{3+4}/\kappa_3$ as a function of $T$ for BAs, AlSb, c-GaN, and w-GaN. Solid curves correspond to calculated $\kappa_{\rm natural}$ and dashed curves correspond to $\kappa_{\rm pure}$.}
\end{figure}

To see more clearly, Fig.\ref{relative_kappa} shows the relative thermal conductivity $\kappa_{3+4}/\kappa_3$ with respect to $T$. The solids curves correspond to the results for materials with naturally occurring isotopic abundances, and dashed curves are for isotopically pure materials. By contrast, it can be concluded that four-phonon scattering is more significant in terms of reducing thermal conductivity for isotopically pure materials than for naturally occurring materials. For AlSb in particular, the RT thermal conductivity of isotopically pure material decreases by more than 70$\%$, much larger than 50$\%$ reduction in naturally occurring case.

\begin{table*}[htbp]
	\centering
	\caption{\label{table:kappa} The calculated RT $\kappa_{\rm natural, 3}$, $\kappa_{\rm natural, 3+4}$, $\kappa_{\rm pure, 3}$, $\kappa_{\rm pure, 3+4}$, and percent isotope effect $P$ for AlSb, BAs, c-GaN, and w-GaN.}
	
	%\begin{tabular}{lcc}
	\begin{tabular}{p{2cm}|p{2cm}<{\centering} p{2cm}<{\centering}| p{2cm}<{\centering} p{2cm}<{\centering} | p{2cm}<{\centering} p{2cm}<{\centering}}
		\hline
		\toprule
		          & \multicolumn{2}{c}{$\kappa_{\rm pure}$ (W/mK)}  \vline & \multicolumn{2}{c}{$\kappa_{\rm natural}$ (W/mK)} \vline & \multicolumn{2}{c}{$P(\%)$}  \\ \hline
		Material & $\kappa_{\rm pure, 3}$   & $\kappa_{\rm pure, 3+4}$ & $\kappa_{\rm natural, 3}$ & $\kappa_{\rm natural, 3+4}$  & $P_3$  & $P_4$ \\ \hline
		AlSb              &181         &51            &99               &50   &83.3  &3.2\\
		BAs               &3322        &1721         &2276         &1441  &46.0   &19.4\\
        c-GaN             &372         &304            &246      &222  & 51.2  &37.2\\
	    w-GaN,a-axis    &406         &359           &228       &223  &78.2   &61.1\\
         w-GaN,c-axis    &450         &385           &261       &243  &72.2   &58.6 \\ \hline \hline
	\end{tabular}
\end{table*}

\begin{figure*}[hbtp]
	\centerline{\includegraphics[width=14cm]{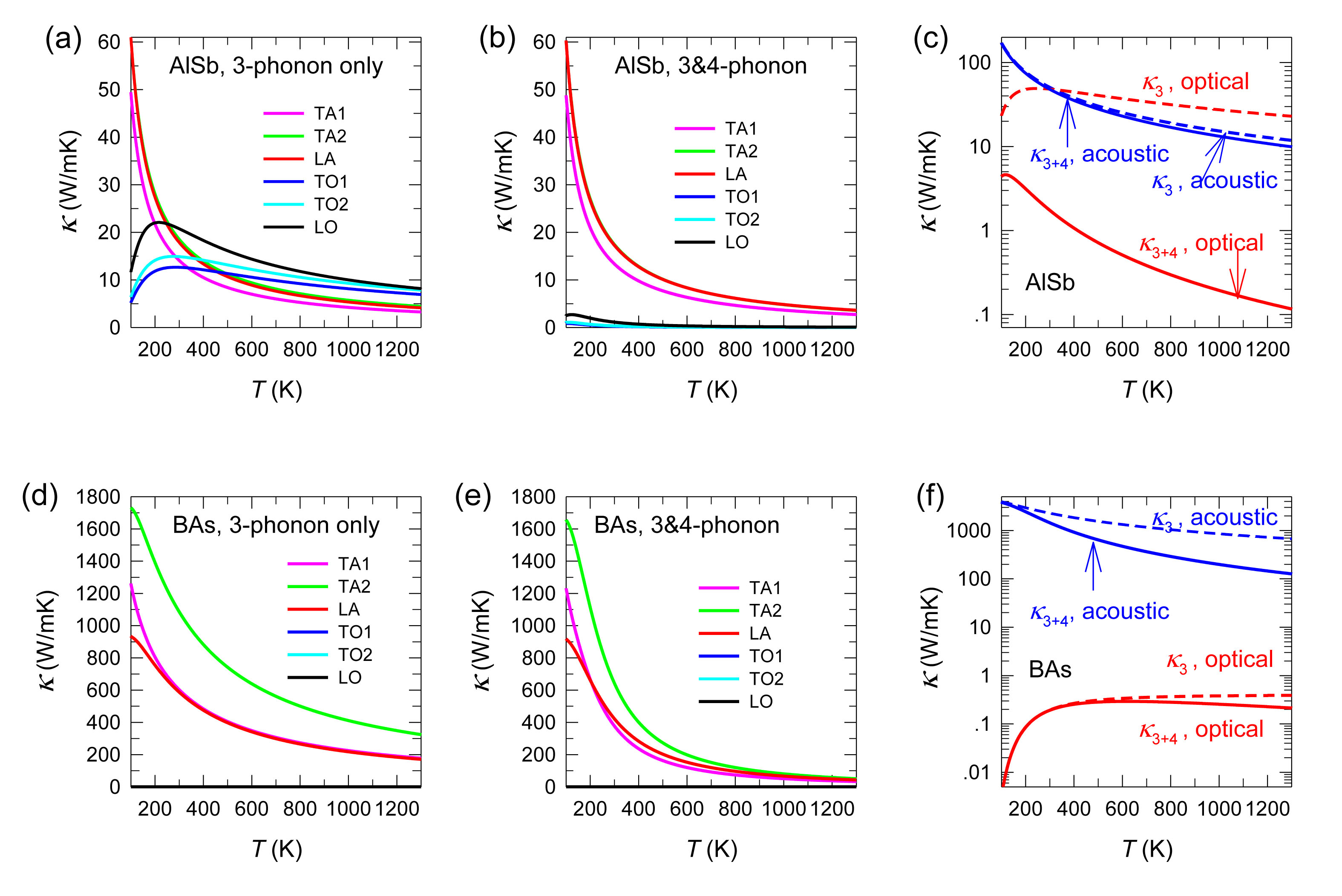}}
	\caption{\label{fig:3}
		$T$-dependent $\kappa_{\rm natural}$ of each branch for AlSb and BAs with $\tau_4^{-1}$ (b, e), and without $\tau_4^{-1}$ (a, d). The contribution to $\kappa_{\rm natural}$ from
		total acoustic branches and total optical branches for AlSb (c) and BAs (f).}
\end{figure*}

To gain further insight into the giant $\kappa$ reduction in AlSb induced by four-phonon scattering, we display the contribution of different phonon branches to $\kappa_{\rm natural}$ with respect to $T$ as shown in Fig.\ref{fig:3} and Table \ref{table:1}. We clearly see from Fig.\ref{fig:3}(a) that with only 3-phonon included, in AlSb the optical branches dominates the $\kappa$ over the acoustic branches even at 250K, and more so at higher temperatures. Surprisingly, as can be seen from Fig.\ref{fig:3}(b), after including four-phonon scattering, not only the acoustic phonon contribution decreases to some extent,
but more importantly, the contribution of all optical modes to the $\kappa$ almost vanishes. For a clearer insight we plot the respective contribution to $\kappa$ from total acoustic modes (blue lines)
and total optical modes (red lines) with increasing $T$ in Fig.\ref{fig:3}(c). We see that in AlSb four-phonon scattering has a much larger impact on optical phonons than acoustic phonons in reducing $\kappa$, while BAs exhibits the opposite pattern (see Fig.\ref{fig:3}(f)). Specifically, after including four-phonon scattering, in AlSb the acoustic contribution to $\kappa$ at RT makes little change, whereas the counterpart of optical contribution is almost fully eliminated. In contrast, in BAs four-phonon scattering leads to $\sim$ 37$\%$ reduction in acoustic contribution and it can be neglected for optical modes which have negligible contribution to the $\kappa$.

In Table \ref{table:1}, we note that in AlSb the relative contributions of all optical branches at 300K
decreases from 48.9$\%$ to 3.6$\%$ after including four-phonon scattering. The contribution of all acoustic branches increases from 51.1$\%$
to 96.4$\%$. Even when $T$ increases to 1000K, the relative contribution of optical modes accounts for only $\sim$ 2$\%$.
This suggests that previous prediction based on three-phonon theory,
showing that optical phonons contributed considerably to the $\kappa$ of AlSb, is actually misleading,
and four-phonon scattering are telling us the fact that the $\kappa$ of AlSb is still primarily
governed by acoustic phonons, and optical phonons have little direct contribution to $\kappa$.
In the case of BAs, unlike AlSb, one can see from Fig.\ref{fig:3}(d-f) whether or not the four-phonon scattering is included,
the acoustic phonons always dominate the $\kappa$.

\begin{table*}[htbp]
	\centering
	\caption{\label{table:1}Percentage contribution of different phonon branches to $\kappa$ for natural occurring AlSb and BAs at 300K and 1000K, respectively.}
	%\begin{tabular}{lcc}
	\begin{tabular}{p{2cm}p{2cm}|p{2.5cm}<{\centering} p{2.5cm}<{\centering}| p{2.5cm}<{\centering} p{2.5cm}<{\centering}}
		\hline
		\toprule
		&          & \multicolumn{2}{c}{300K}  \vline & \multicolumn{2}{c}{1000K} \\ \hline
		Material & $\kappa \%$  & 3-phonon only  & 3$\&$4-phonon & 3-phonon only  & 3$\&$4-phonon \\ \hline
		AlSb    & TA1          &14.26$\%$         &26.91$\%$            &9.89$\%$               &27.05$\%$ \\
		& TA2          &18.77$\%$         &34.65$\%$            &13.42$\%$           &34.09$\%$ \\
		& LA           &18.09$\%$         &34.83$\%$            &12.52$\%$           &36.78$\%$ \\
		& Optical      &48.88$\%$         &3.61$\%$            &64.17$\%$           &2.08$\%$ \\ \hline
		BAs     & TA1          &28.35$\%$         &27.46$\%$            &28.27$\%$         &26.52$\%$  \\
		& TA2          &47.64$\%$         &44.77$\%$           &47.94$\%$          &40.86$\%$ \\
		& LA           &24$\%$            &27.75$\%$           &23.73$\%$          &32.45$\%$ \\
		& Optical      &0.01$\%$         &0.02$\%$           &0.06$\%$               &1.7$\%$ \\ \hline \hline
	\end{tabular}
\end{table*}

\begin{figure}[hbtp]
	\centerline{\includegraphics[width=8cm]{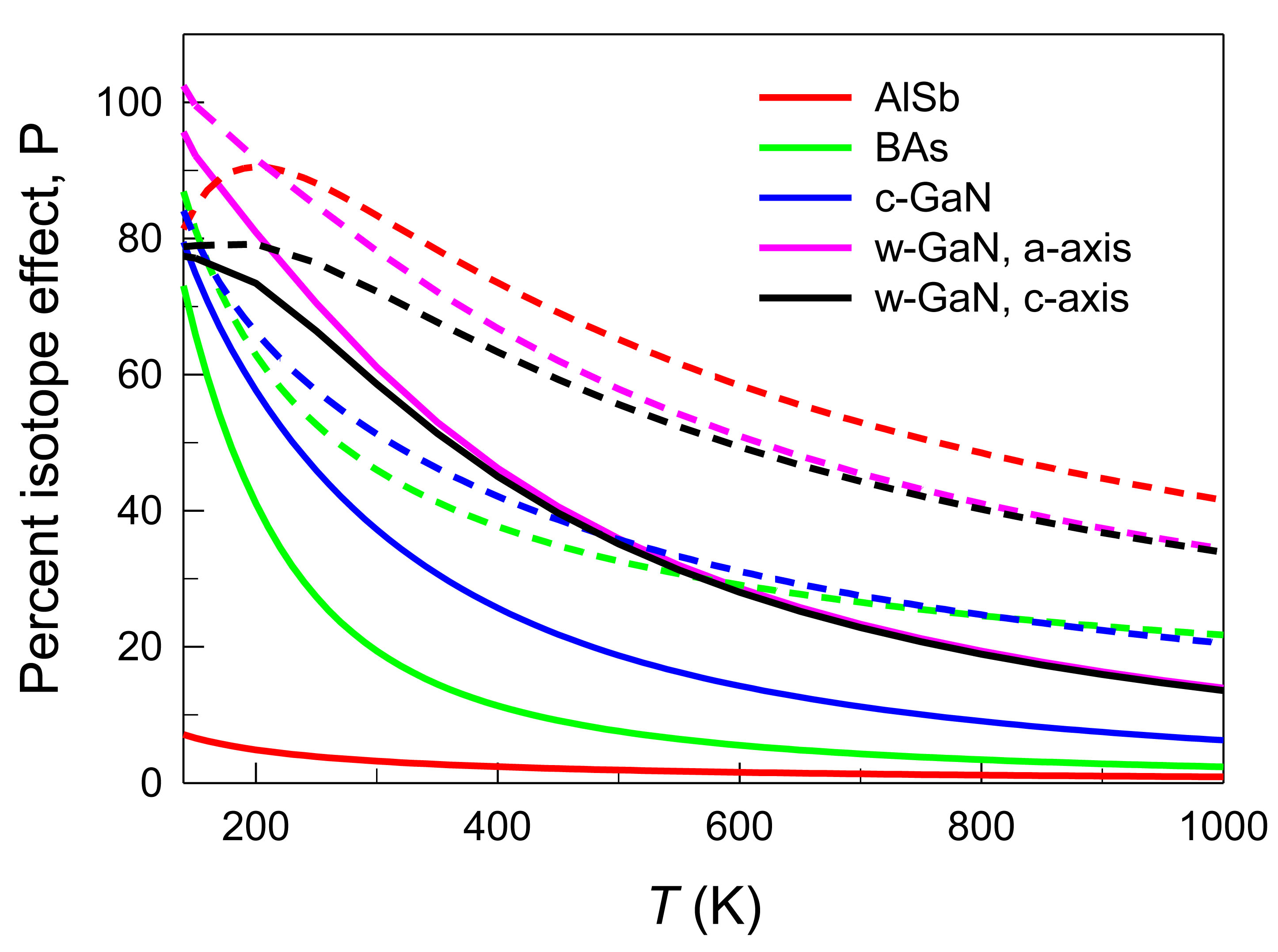}}
	\caption{\label{Peffect}
    Calculated isotope effect, $P = (\kappa_{\rm pure}/\kappa_{\rm natural}-1)\times100\%$, versus temperature for AlSb(red), BAs
(blue), c-GaN(pink), and w-GaN (green). Dashed lines give the results with only three-phonon scattering, and solid lines give the results after including four-phonon scattering.}
\end{figure}

We then show the calculated isotope effect, defined by $P = (\kappa_{\rm pure}/\kappa_{\rm natural}-1)\times100\%$, versus $T$ for AlSb (red lines), BAs (green lines), c-GaN (blue lines), and w-GaN (pink and black lines) in Fig.\ref{Peffect}. Dashed lines give the results with only three-phonon scattering, and solid lines give the results after including four-phonon scattering. Our calculated RT $P$ for these materials are also given in Table \ref{table:kappa}. Comparing AlSb, BAs, c-GaN, and w-GaN, we find that the isotope effect $P$ is significantly weakened by four-phonon scattering, especially for AlSb. With only three-phonon scattering included, the largest $P$ values occur for AlSb, with $P=83.3\%$ at RT, and the calculated RT $P$ values for BAs, c-GaN, and w-GaN are $46.0\%$, $51.2\%$ and $72.2\sim 78.2\%$, respectively. Also, we note that for AlSb a peak in $P$ occurs at around 200K, which is attributed to the interplay between phonon-phonon and phonon-isotope scattering as mentioned before. With four-phonon scattering included, interestingly, we find that AlSb has the smallest $P$ value among these materials, and its RT $P$ decreases to $3.2\%$. In contrast, w-GaN has the highest calculated $P$ among all the materials with $P=56.8\sim 61.1\%$ at RT, and the next highest calculated $P$ is that of c-GaN, with $P=37.2\%$ at RT, and the calculated RT $P$ for BAs is $P=19.4\%$. These results are intriguing since four-phonon scattering can greatly weaken the isotope effect on $\kappa$ for these materials with the large a-o gap, where the phonon-isotope scattering plays a significant or even dominant role over the anharmonic three-phonon scattering in determining $\kappa$.

\begin{figure*}[hbtp]
	\centerline{\includegraphics[width=20cm]{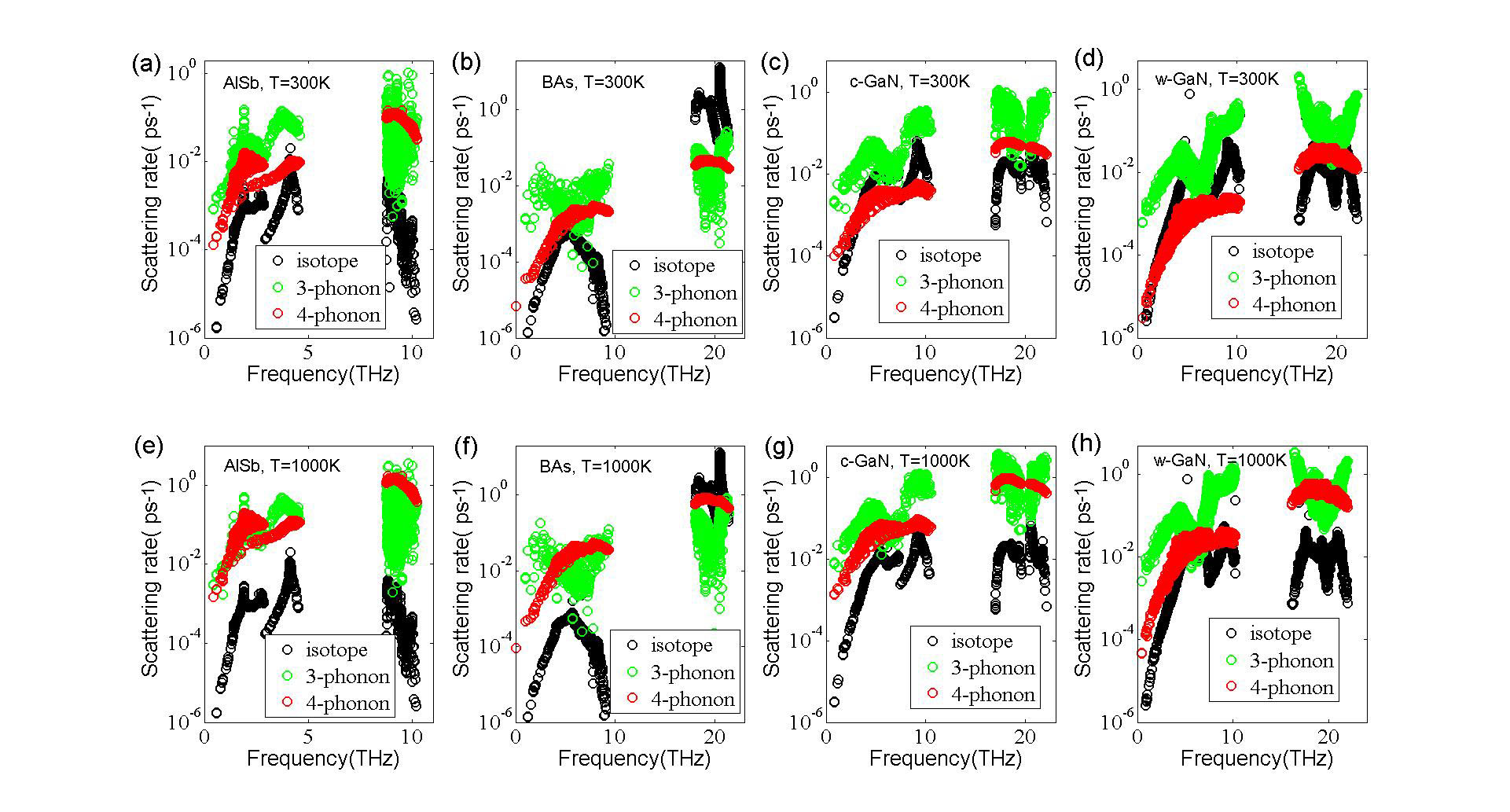}}
	\caption{\label{SRs}
   Calculated intrinsic three-phonon scattering rates (blue circles), four-phonon scattering rates (red circles), and the isotope scattering rates (black circles) as a function of frequency for AlSb (a, e), for BAs (b, f), for c-GaN (c, g), and for w-GaN (d, h) at 300 and 1000K, respectively.}
\end{figure*}

To understand the above results better, Figs.\ref{SRs} give the intrinsic three-phonon scattering rates $\tau_3^{-1}$, four-phonon scattering rates $\tau_4^{-1}$, and the phonon-isotope scattering rates $\tau_{\rm iso}^{-1}$ as a function of frequency for these four materials studied. We find that for AlSb and BAs the $\tau_4^{-1}$ is very strong and comparable to or even much higher than $\tau_3^{-1}$ even at RT, and more so at 1000K, especially for optical phonons, while $\tau_{\rm iso}^{-1}$ is comparable to $\tau_3^{-1}$ at RT or much lower temperatures (see Figs.\ref{SRs} (a, b)). As $T$ increases to 1000K, the isotope scattering becomes relatively less important as compared to anharmonic phonon-phonon scattering, as shown in Figs.\ref{SRs}(e, f). Since the intrinsic phonon-phonon scattering rates scale with $T$ as $AT+BT^2$ whereas the isotope scattering rates are $T$ independent, causing that the isotope effect becomes less important with increasing $T$. Remarkably, we can also see from Fig.\ref{SRs} (b, f) that in BAs the isotope scattering rates of optical modes are much larger than the three- and four-phonon scattering even at 1000K, indicating that the isotope effect plays a dominant role in infrared optical phonon linewidths of BAs. As for c-GaN and w-GaN, as is seen from Fig.\ref{SRs} (c, d), the four-phonon scattering rates are much weak than the three-phonon and isotope scattering rates at RT, but still cannot be neglected. When $T$ goes up to 1000K, we find that for both c-GaN and w-GaN the four-phonon scattering rates are comparable to the three-phonon scattering rates, as shown in Fig.\ref{SRs}(g, h). These results clearly demonstrate the importance of four-phonon scattering in determining $\kappa$ for both natural occurring materials and isotopically pure materials with a large band gap.

\begin{figure*}[hbtp]
	\centerline{\includegraphics[width=12cm]{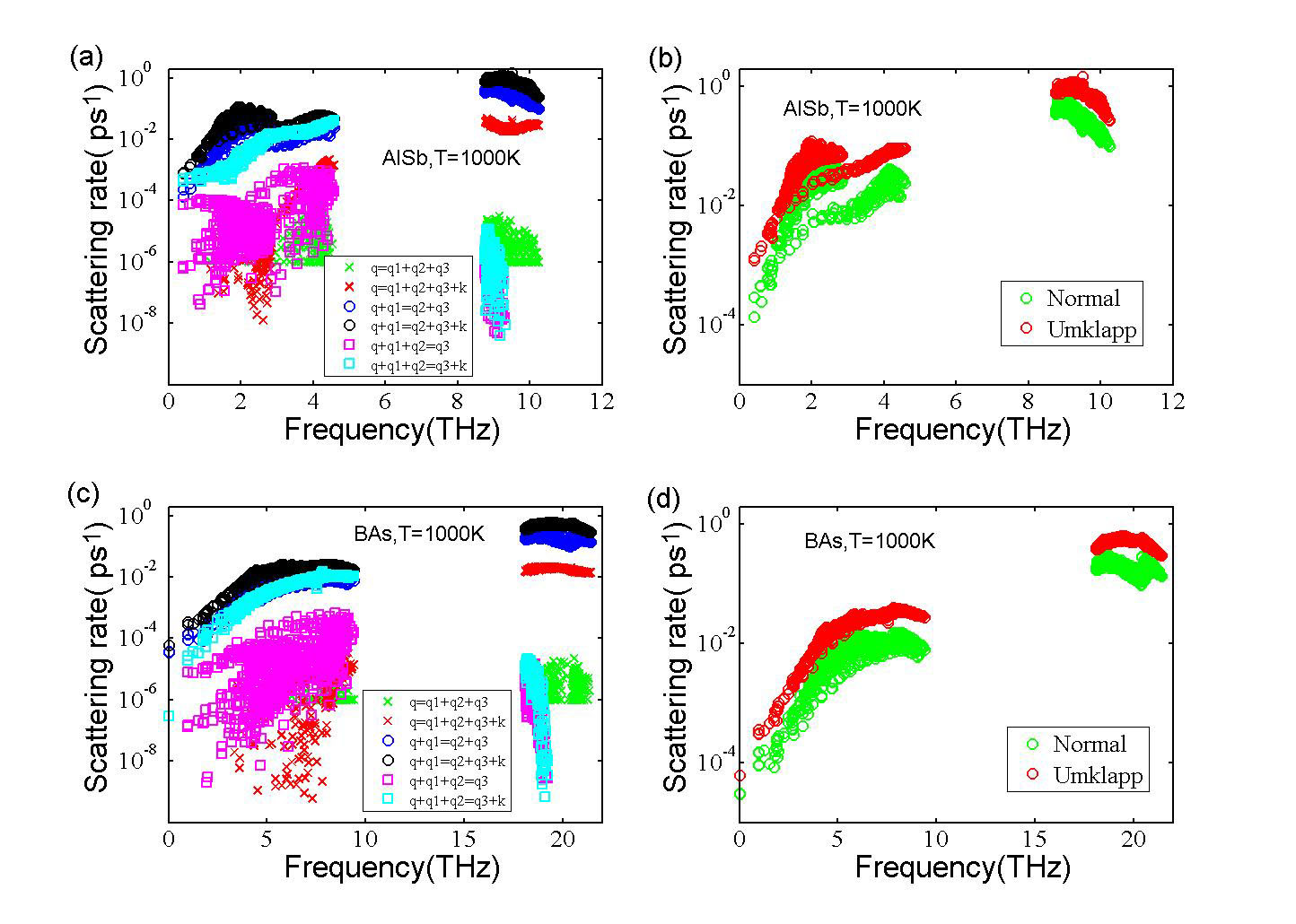}}
	\caption{\label{fig:5}
 The contribution to $\tau_4^{-1}$ from all allowed scattering channels as a function of frequency for AlSb (a), and BAs (c) at 1000K. The contributions of Umklapp processes to $\tau_4^{-1}$ are compared to
 the contribution of Normal processes for AlSb (b), and BAs (d) at 1000K. }
\end{figure*}

To gain deeper understanding of four-phonon scattering, taking BAs and AlSb as examples, we further study the contribution to four-phonon scattering rates from all possible scattering channels.
Clearly, we see from Figs.\ref{fig:5}(a, c) that for both BAs and AlSb, the dominant scattering channels for acousitc modes are recombination processes
$\textbf{q}+\textbf{q}_1 \longrightarrow \textbf{q}_2+\textbf{q}_3+\textbf{k}$ and $\textbf{q}+\textbf{q}_1 \longrightarrow \textbf{q}_2+\textbf{q}_3$, and absorption process
$\textbf{q}+\textbf{q}_1+$\textbf{q}$_2 \longrightarrow \textbf{q}_3+\textbf{k}$, where $\textbf{k}$ is a reciprocal lattice vector, which is zero for Normal processes and non-zero for
Umklapp processes. While the dominant scattering channels for optical modes are splitting process
$\textbf{q} \longrightarrow \textbf{q}_1+$\textbf{q}$_2+\textbf{q}_3+\textbf{k}$, and recombination processes
$\textbf{q}+\textbf{q}_1 \longrightarrow \textbf{q}_2+\textbf{q}_3+\textbf{k}$ and $\textbf{q}+\textbf{q}_1 \longrightarrow \textbf{q}_2+\textbf{q}_3$.
Amongst these scattering processes, the recombination process always makes a major contribution to $\tau_4^{-1}$, whether acoustic or optical phonons. Due to the bunching of optical branches, these processes can easily satisfy the energy and momentum conservation rules, as illustrated in Fig.\ref{band}, unlike the absorption or splitting process, which strongly depends on the $T$ and a-o gap.
Also, we find that for both systems four-phonon scattering is dominated by resistive Umklapp processes as shown in Figs.\ref{fig:5}(b) and (d), which also demonstrates the effectiveness of our iterative method described above.

\begin{table*}[htbp]
\centering
\caption{\label{table:2}
Calculated Debye temperature $\theta_{\rm{D}}$, a-o gap $E_{\rm{g}}$, highest LA phonon frequency, normalized a-o gap, and relative thermal conductivity
$\kappa_{3+4}/\kappa_3$ at RT and $T=\theta_{\rm{D}}$ in AlSb, BAs, c-GaN and w-GaN.}

%\begin{tabular}{lcc}
\begin{tabular}{p{3cm}<{\centering} p{2cm}<{\centering} p{2.5cm}<{\centering} p{2cm}<{\centering}p{2cm}<{\centering}p{2.5cm}<{\centering}p{3.2cm}<{\centering}}
\hline
\toprule
Material  & $\theta_{\rm{D}}$ [K]   & $E_{\rm{g}}$ [THz]  & $\omega_{\rm{LA}}$ [THz] & $E_{\rm{g}}/\omega_{\rm{LA}}$   & RT-$\kappa_{3+4}/\kappa_3$ [$\%$] & $\kappa_{3+4}/\kappa_3$ ($T=\theta_{\rm{D}}$) [$\%$]  \\ \hline
AlSb             & 276                     & 4.3                         & 4.5                      & 0.96                                 & 50.3                              & 52.6  \\
BAs              & 651                     & 9.1                         & 9.6                      & 0.95                                 & 63.3                              & 33.2  \\
c-GaN            & 584                     & 6.0                         & 10.6                     & 0.57                                 & 90.1                              & 74.4  \\
w-GaN (a-axis)   & 584                     & 6.1                         & 9.4                      & 0.65                                 & 97.7                              & 88.7  \\ 
w-GaN (c-axis)   & 584                     & 6.1                         & 9.4                      & 0.65                                 & 93.0                              &84.7 \\ \hline \hline
\end{tabular}
\end{table*}

Moreover, to generalize the impact of four-phonon scattering on the $\kappa$ of any solids, we normalize the a-o gap $E_{\rm{g}}$ by the highest LA phonon frequency $\omega_{\rm{LA}}$
for ease of cross-comparing between different materials.
Table\ref{table:2} lists the calculated Debye temperature $\theta_{\rm{D}}$, $E_{\rm{g}}$, $\omega_{\rm{LA}}$, normalized gap $E_{\rm{g}}/\omega_{\rm{LA}}$, and relative thermal conductivity
$\kappa_{3+4}/\kappa_3$ at RT and $\theta_{\rm{D}}$ for materials studied in this work.
Comparing AlSb, BAs and c-GaN, which have the same crystal structure, we find that RT-$\kappa_{3+4}/\kappa_3$ increases monotonically with decreasing $E_{\rm{g}}/\omega_{\rm{LA}}$, since the larger a-o gap, the weaker three-phonon scattering involving a-o coupling, and thus the more important four-phonon scattering, as demonstrated previously \cite{Feng2017}. At $T=\theta_{\rm{D}}$, however, $\kappa_{3+4}/\kappa_3$ does not change monotonically with the band gap.
Typically, although $E_{\rm{g}}/\omega_{\rm{LA}}$ is larger in AlSb than in BAs, $\kappa_{3+4}/\kappa_3$ of BAs at $T=\theta_{\rm{D}}$ is obviously smaller.
The reason for this is that for BAs, besides the large a-o gap, the bunching of acoustic band also restrict  three-phonon interaction as mentioned above, which however, cannot affect four-phonon processes.
Also, by comparing between c-GaN and w-GaN, we find that although they have almost same band gap, in c-GaN the $\kappa$ with $\tau_4^{-1}$ decreases by 25.6$\%$ at $T=\theta_{\rm{D}}$, while the reduction of $\kappa$ induced by $\tau_4^{-1}$
is less than 14$\%$ in the case of w-GaN. This can be explained from the difference in phonon dispersions of both materials. We notice that comparing to c-GaN,
in w-GaN there exists a few optical branches with frequency below acoustic limit, as shown in Fig.\ref{band}. Although these modes do not significantly contribute directly to $\kappa$, they can provide important scattering channels for the heat-carrying acoustic phonons, where three-phonon process has a much higher scattering probability than that of four-phonon process, thus showing that in this case, the effect of four-phonon scattering will shrink. Our analysis above indicates whether four-phonon scatttering is important not just depends on the a-o gap, which is actually connected to the interplay among the a-o gap, the acousitc bunching, and the number of optical phonons with low frequency.

To summarize, we employ a first-principles approach to revisit the $\kappa$ and the isotope effect in AlSb, c-GaN, and w-GaN in comparison to BAs, by including four-phonon scattering. We demonstrate that in AlSb where three-phonon scattering involving optical phonons is extremely weak due to the unusual dispersion features (i.e. large a-o gap, flatness of optical bands), four-phonon scattering can actually play a dominant role in the optical phonon thermal resistance. Surprisingly, we find that four-phonon scattering completely diminishes optical phonon thermal conductivity, and leads to $\sim$ 50$\%$ reduction in the total $\kappa$ of natural occurring AlSb even at RT, which is even larger than that found in BAs. We believe this finding is not a particular case, and it is expected to widely exist in other semiconductor materials, which have similar phonon features with AlSb. Also, our calculation results show that four-phonon scattering can play an extremely important role in weakening the isotope effect on $\kappa$, especially for these materials with unusually weak anharmonic three-phonon scattering. Typically, for AlSb when four-phonon scattering is added, the isotope effect $P$ at RT will be reduced significantly, from $83.2\%$ to $3.2\%$.
Furthermore, we demonstrate the general importance of four-phonon scattering in the intrinsic $\kappa$ of materials, which is related to the interplay among the large a-o gap, acoustic bunching, and low-frequency optical branches. Our work gives a critical revisit to exactly predicting the $\kappa$ values of materials where optical phonons contribute considerably, and also provides the rigorous understanding of the interplay between intrinsic phonon-phonon scattering and isotopic scattering.

%\clearpage
%\bibliography{refs}

\end{document}